\documentclass[preprint]{aastex}

\shorttitle{MODE CONVERSION OF AN EUV WAVE}
\shortauthors{Zong \& Dai}

\begin{document}

\title{MODE CONVERSION OF A SOLAR EXTREME-ULTRAVIOLET WAVE OVER A CORONAL CAVITY}

\author{Weiguo Zong\altaffilmark{1,3}, Yu Dai\altaffilmark{2,3}}
\altaffiltext{1}{Key Laboratory of Space Weather, National Center for Space Weather, China Meteorological Administration, Beijing 100081, China}
\altaffiltext{2}{School of Astronomy and Space Science, Nanjing University, Nanjing 210023, China}
\altaffiltext{3}{Key Laboratory of Modern Astronomy and Astrophysics (Nanjing University),
Ministry of Education, Nanjing 210023, China}

\email{ydai@nju.edu.cn}

\begin{abstract}
We report on observations of an extreme-ultraviolet (EUV) wave event in the Sun on 2011 January 13 by \emph{Solar Terrestrial Relations Observatory} (\emph{STEREO}) and \emph{Solar Dynamics Observatory} (\emph{SDO}) in quadrature. Both the trailing edge and the leading edge of the EUV wave front in the north direction are reliably traced, revealing generally compatible propagation velocities in both perspectives and a velocity ratio of about 1/3. When the wave front encounters a coronal cavity near the northern polar coronal hole, the trailing edge of the front stops while its leading edge just shows a small gap and extends over the cavity, meanwhile getting significantly decelerated but intensified. We propose that the trailing edge and the leading edge of the northward propagating wave front correspond to a non-wave coronal mass ejection (CME) component and a fast-mode magnetohydrodynamic (MHD) wave component, respectively. The interaction of the fast-mode wave and the coronal cavity may involve a mode conversion process, through which part of the fast-mode wave is converted to a slow-mode wave that is trapped along the magnetic field lines. This scenario can reasonably account for the unusual behavior of the wave front over the coronal cavity.
\end{abstract}


\maketitle

\section{INTRODUCTION}
First discovered by the Extreme-Ultraviolet (EUV) Imaging Telescope (EIT) on board the \emph{Solar and Heliospheric Observatory} (\emph{SOHO}), the so-called ``EIT waves" occurring during solar eruptions such as flares and coronal mass ejections (CMEs), are characterized by a diffuse bright front globally propagating  through the solar corona \citep{mos97,Thom98}. EIT waves were initially interpreted as a fast-mode magnetohydrodynamic (MHD) wave in the corona \citep{Thom99}, whose base, if strong enough, can also perturb the chromosphere to produce an apparently fast-propagating H$\alpha$ Moreton wave \citep{moren60}, as modeled in \citet{uchi68}.  Nevertheless, alternative models were also proposed that regard EIT waves as a result of magnetic reconfiguration related to the CME liftoff rather than a true wave. Such non-wave models include the current sheet model \citep{dela00}, the field-line stretching model \citep{chen02,chen05}, and the successive reconnection model \citep{att07}.

With observations from modern generation of EUV imaging instruments on board the 
 \emph{Solar Terrestrial Relations Observatory} \citep[\emph{STEREO};][]{kai08} and \emph{Solar Dynamics Observatory} \citep[\emph{SDO};][]{Pesnell12}, now we prefer the term ``EUV waves" to the conventional ``EIT waves". New observations of EUV waves with the unprecedentedly high temporal and spatial resolutions often reveal the co-existence of a fast-propagating front ahead of a slow-propagating one \citep[etc.]{liu10,chen11,liu12,cheng12}. The slow front is found co-spatial with the CME frontal loop \citep{dai12}, while the fast front shows wave characteristics such as reflection from a coronal hole or a coronal bright structure \citep{gop09, Li12, Ole12}, diffraction or refraction upon the interaction with a remote active region \citep{shen13}, and triggering of remote filament oscillations \citep{dai12, zong15}. These observations reasonably incorporate both the wave and non-wave components into an individual EUV wave event.

Early observations showed that an EUV wave would stop at a magnetic separatrix or quasi-separatrix layer \citep[QSL;][]{dela99}, which has been regarded as a compelling argument against the wave model. Recently, \citet{chan16} reported a new feature in an EUV wave that the fast-mode wave passes through a QSL, leaving a stationary bright front behind. Motivated by this founding, \citet{chen16} numerically simulated the passage of a fast-mode shock wave through a QSL. The results show that the interaction of the fast-mode wave and the QSL results in a stationary wave front in front of the QSL. They proposed that part of the fast-mode may be converted to a slow-mode wave, which is then trapped inside the magnetic loops. Up to now, unambiguous observations of mode conversion in an EUV wave have not yet been reported. In this Letter, we present quadrature observations of an EUV wave event and its interaction with a coronal cavity, which, to our knowledge, for the first time provide observational evidence of mode conversion in an EUV wave event.

\section{OBSERVATIONS AND DATA ANALYSIS}
The EUV wave event was launched on 2011 January 13 from NOAA AR 11147.  The position information of the AR and the \emph{STEREO} twin spacecraft on the date of the event is depicted in Figure 1(b), revealing a quadrature configuration of both the spacecraft with respect to the Earth and a location of the AR near the central meridian in the field of view (FOV) of \emph{STEREO Behind}  (\emph{STB}). Although the AR was still behind the eastern solar limb from the Earth perspective  (e.g. \emph{SDO}), the occultation effect was not so significant that substantially it blocked our view on the  off-limb structures of the EUV wave.  Therefore, in this work, we utilized quadrature EUV imaging observations from the EUV Imager (EUVI) on board  \emph{STB} (hereafter EUVI-B) and the Atmospheric Imaging Assembly (AIA) onboard \emph{SDO} to trace the evolution of the EUV wave on-disk and off-limb, respectively.  During the interval of interest, the EUVI-B 195~{\AA} channel, which best captures EUV waves, took images with a time cadence of 2.5 minutes and a pixel size of $\sim$1$\farcs$6. AIA  provides multiple simultaneous full-disk images of the transition region and corona in 10 EUV and UV channels with 0$\farcs$6 pixel size and 12 s temporal resolution. Besides, the Large Angle and Spectrometric COronagraph \citep[LASCO;][]{bru95} C2 instrument on board \emph{SOHO} traces CMEs from 2 to 6~$R_\odot$ with a pixel size of 11$\farcs$9. Spatial comparison between the CME structure and the associated EUV wave front in the low corona can help us better understand the nature of the event.

Figure 1 shows the pre-event coronal environment from different viewpoints and in different channels. In EUVI-B 195 {\AA} (Figure 1(a)),  the core of the AR  is composed of compact low-lying bright loops.  Fainter loops rooted in the peripherals of the AR extend to higher altitudes, being also observable in AIA. To the north of the AR, there is a long dark channel near the northern polar coronal hole. In AIA (Figures 1(c-f)), the cross section of this structure is identified by a spike-like prominence (in 304 and 171 {\AA}) surrounded  by a dark cavity (in 193 and 211 {\AA}) over the limb, which conforms to the observational criteria for a coronal cavity used in \citet{karna15}. By carefully checking the date and the location of the coronal cavity with the online cavity catalog (\url{http://spaceweather.gmu.edu/projects/synop}), we found that this cavity is entitled as C210558133, which means that it appeared during Carrington rotation 2105 with an average latitude of 58$^{\circ}$ and average longitude of 133$^{\circ}$.

The evolution of the eruption and the associated EUV wave is presented in the online animations. Figure 2 shows some snapshots of the early stage evolution nearly simultaneously in EUVI-B 195 {\AA} and AIA 193 {\AA}\@. In the beginning, erupting structures are seen as bright sharp loops ascending from the AR core (Figures 2(a) and (e)).  By using the triangulation method, we could identify the same feature in an erupting loop seen from the two perspectives, which is outlined by the plus signs.  It is seen that thanks to the low off-limb background, some more diffuse loops are observed stretched ahead of the erupting loops in AIA, which are barely discernible against the solar disk in EUVI-B\@.  As the eruption evolves, the top parts of the stretched loops gradually fade out and leave the FOV of AIA, while the footpoints of these loops  brighten consecutively along the limb (Figures 2(f-h)). In EUVI-B, these brightenings form a quasi-circular diffuse EUV wave front (outlined by the arrows) that propagates outward from the AR, leaving dimmings behind (Figures 2(b-d)). As seen from AIA, the erupting loops do not ascend in the purely radial direction but somewhat inclined to the north. As a result, the EUV wave is more prominent in the north direction.

To explore the full evolution of the EUV wave, we used time--distance stack plots to study the wave kinematics. In EUVI-B 195 {\AA}, we selected two sectors  labeled as  ``A'' and ``B'' in Figure 1(a). The two sectors extend an angular span of 10$^{\circ}$ from the AR core to an area of relatively quiet Sun to the south (A),  and across the dark channel (coronal cavity) to the north (B), respectively. In constructing the time--distance stack plot for each sector, the perturbation profile at every observing time was derived by averaging the base ratio intensity values in annuli of 1$^{\circ}$ width with increasing radii on the spherical solar surface. The resulting stack plots are shown in Figures 3(a-b). By applying linear fits to the edges of moving structures on the wave fronts,  the wave velocities in different directions were obtained. In sector A,  the EUV wave propagates roughly continuously over the largely quiet Sun region at a velocity of $\sim$240 km s$^{-1}$.  In sector B,  both the leading edge and the trailing edge of the wave front that encloses the dimming region can be reliably traced; they reveal velocities of $\sim$290 km s$^{-1}$ and $\sim$90 km s$^{-1}$, respectively, leading to an increasingly broadening front.  While the trailing edge of the front finally stops near the southern boundary of the cavity, the propagation of  the leading edge just shows a small gap at there and extends over the cavity. Meanwhile, the amplitude of the wave north to the cavity is obviously enhanced (seen from the color bar). Interestingly,  tracing the most enhanced features of the front ($\sim$20\% over the pre-event level) turns out an apparent backward propagation from the northern polar coronal hole at a velocity of $\sim$190 km s$^{-1}$.

In AIA, the EUV wave is found to propagate along the eastern solar limb. To avoid any ambiguities introduced from close-to-limb disk regions, we studied the off-limb wave behavior along a semi-circle of 10 AIA pixels ($\sim$6 arcsec) in thickness at a height of 0.1 $R_{\sun}$ above the eastern solar limb (shown in Figure 1(c)). By using a similar method to that applied to the sectors in EUVI-B, we generated the time--position angle (PA) stack plots of the wave front along the semi-circle in AIA 171, 193, and 211 {\AA} channels, respectively, as shown in Figures 3(d-f). As found in \citet{dai12}, the main body of the EUV wave front appears darkening in the cooler 171 {\AA} channel, but moderately and strongly brightening in the medium 193 {\AA} and warmer 211 {\AA} channels , indicative of a heating process within the wave front. Following the front are dimmings in 193 and 211 {\AA} but brightening structures in 171 {\AA}\@. By checking the online animation, it is found that the brightening structures in 171 {\AA} are composed of many stretched loops, which may have cooled down from a high temperature. Despite the distinct emission behaviors, the kinematics of the wave front is almost the same among different channels. To the south, the front moves laterally at a velocity of $\sim$290 km s$^{-1}$, while to the north, the propagation velocities of the leading and trailing edges of the front are $\sim$380 km s$^{-1}$ and $\sim$120 km s$^{-1}$, respectively. The trailing edge that encloses the dimmings in 193 and 211 {\AA} (or brightenings in 171 {\AA}) stops at a PA of 56$^{\circ}$, corresponding to the southern boundary of the cavity. For the leading edge, its propagation is also somewhat interrupted by the cavity and the front reappears beyond a PA of 67$^{\circ}$ (northern boundary of the cavity), meanwhile getting more brightened in 211 {\AA}. Nevertheless, the front turns from originally brightening to darkening in 193 {\AA} while in 171 {\AA} it keeps darkening. A linear fit to this front reveals a much lower forward propagation velocity of $\sim$160 km s$^{-1}$  before it finally stops at a PA of 78$^{\circ}$, near the boundary of the northern polar coronal hole.

After compensating for the difference in tracing heights, the wave velocities obtained from EUVI-B and AIA are generally compatible but still show discrepancies of 30--60 km s$^{-1}$. The reason for the discrepancies may be that since the source AR is located behind the limb in AIA, from the two different perspectives we are unable to trace exactly the same features, which further reflects a 3D geometry of the wave structure.

The eruption finally develops to a CME into the interplanetary space. In Figure 4 the composite images of LASCO C2 and AIA in 211 {\AA}  demonstrate the evolution of the CME  and  the EUV wave in a combination.  We focused on the northern flank of the CME for the reasons that (1) the propagations of both the CME and the EUV wave are biased to the north direction, and (2) in this direction there is an interesting phenomenon that the wave front interacts with the coronal cavity and then gets intensified. As mentioned above, upon the first appearance of the CME in the FOV of LASCO C2 (around 09:36 UT), the leading edge of the EUV wave front has propagated across the cavity and extended beyond a PA of 67$^{\circ}$, which corresponds to the northernmost PA of the cavity. For the CME northern flank, it is found that its lateral expansion once approaches this PA (09:48 UT and 10:00 UT),  but later on shrinks to a PA  of 56$^{\circ}$ (after 10:12 UT), consistent with the extent of the trailing edge of the EUV wave front that stops at the southern boundary of the cavity. No obvious CME ejecta are observed beyond the PA of 67$^{\circ}$, while the intensified EUV wave front over the cavity can be traced to a farther PA.

\section{DISCUSSION AND CONCLUSIONS}
An EUV wave event occurring on 2011 January 13 was observed with \emph{STB}/EUVI and \emph{SDO}/AIA in quadrature.  It is noted that during early 2009 when the \emph{STEREO} twin spacecraft were in a quadrature configuration, similar quadrature \emph{STEREO} observations of EUV waves with a suitable source region location have been reported in literatures \citep[e.g.,][]{Pats09,Kien09}. Such observations enable us to better understand the 3D structure of an EUV wave.

Combining both the EUVI-B on-disk and the AIA off-limb observations together, it is found that the eruption is initiated as bright sharp loops ascending from the core of the source AR, which subsequently stretch and perturb the overlying magnetic field lines. The perturbation propagates downward along the filed lines, causing brightenings at the footpoints of these lines along the limb, as seen from AIA. When projected onto the disk, these brightenings form a diffuse front propagating away from the AR in EUVI-B\@. This is consistent with the field-line stretching model of EUV waves proposed by \citet{chen02}, in which a slow EUV wave front corresponds to the frontal loop of the driving CME. In their model, the perturbation can also propagate across the field lines, forming another fast-propagating front that is indeed a fast-mode MHD wave. For the front in EUVI-B sector B as well as the northward propagating front along the semi-circle in AIA, it is found that the propagation velocity of the trailing edge of the front (90/120 km s$^{-1}$) is about $1/3$ of that of the leading edge (290/380 km s$^{-1}$), just as the velocity ratio of the two fronts in \citet{chen02} model derived by assuming a semi-circular field-line geometry. Therefore, it is reasonable to attribute the trailing edge to a non-wave CME component and regard the leading edge as a fast-mode wave component. Since the eruption is inclined to the north, the CME frontal loops can continuously drive fast-mode waves in this direction. Hence, the preceding wave component and the following non-wave component in the visible wave front may not be separated from each other while the front is just increasingly broadening. In the south direction, we found from the online animation that the footpoints of the stretched loops are much more converged. Therefore, the fronts seen in EUVI-B sector A and the southward propagating front in AIA are most likely a freely propagating fast-mode wave driven by only a short southward expansion of the CME, thus showing no signatures of wave broadening. Without a continuous driver, the wave front in the south direction should propagate somewhat slower (240/290 km s$^{-1}$) than that in the north direction. Furthermore, both the non-wave CME component and the fast-mode wave component involve adiabatic plasma compression, which can lead to some extent of plasma heating within the wave front. This may explain the distinct emission behaviors of the wave front seen in the different AIA channels.

In the \citet{chen02} model, the non-wave component of an EUV wave should stop at a magnetic separatrix or QSL. According to the definition of coronal cavities in \citet{karna15}, a coronal cavity is an elongated filament channel along the east--west direction, whose boundaries can serve as a good proxy for QSLs. In this work, the trailing edge of the northward propagating front stops at the southern boundary of the cavity, which is also in good agreement with the final PA of the CME northern flank, further consolidating the non-wave nature of this component. On the other hand, its leading edge extends over the cavity and gets intensified. In EUVI-B the intensified front shows an apparent backward propagation, looking like  wave reflection by the polar coronal hole \citep{gop09, Li12, Kien13}, while in AIA the front continues propagating forward but at a much lower velocity (160 km s$^{-1}$) before forming a stationary front near the boundary of the northern polar coronal hole. We think that the observational factors seen in AIA perfectly resemble the numerical results in \citet{chen16}  that the interaction of a fast-mode wave and a QSL results in a stationary wave front in front of the QSL, which they attributed to mode conversion \citep[cf.][]{cal05}. In Figure 5 we schematically show how the mode conversion process can affect the wave behavior over the cavity in this event. As the trailing edge has been stopped by the cavity, the leading edge (believed to be a fast-mode wave) continues its propagation and interacts with the cavity. Through the interaction (depicted by the asterisk), part of the fast-mode wave is converted to a slow-mode (most likely an acoustic-mode) wave, which is then trapped inside the magnetic loops. When tracing the wave front along an arc line, as shown in the AIA time--PA stack plots, we will anticipate a significantly ``decelerated" front over the cavity. For a slow-mode wave along the magnetic field, the perturbation strength only increases the plasma pressure rather than synchronous enhancement of the plasma and magnetic pressure in case of a fast-mode wave perpendicular to the field, thus leading to more plasma compression and then further heating. This may explain why the front gets more brightened in 211 {\AA} but becomes darkening in 193 {\AA} after the mode conversion.  As to the apparent backward propagation of the mode-converted wave front in EUVI-B, we conjecture that this is a line-of-sight (LOS) integration effect rather than true wave reflection. As shown in Figure 5, the slow-mode wave front propagates downward from ``f1" to ``f2" along the field line, while seen along the LOS in EUVI-B the two off-limb fronts are projected onto the solar surface. The cavity is located at a high latitude  where the solar surface is quite tangential to the LOS in EUVI-B\@. As long as the field line is not very parallel to the LOS, in EUVI-B we will see a backward propagating front against the solar disk.

\begin{acknowledgements}
We are grateful of the anonymous referee for his/her suggestive comments. The authors thank the \emph{STEREO}/SECCHI, \emph{SDO}/AIA, and \emph{SOHO}/LASCO consortia for their open data policy.  This work is supported by NSFC (40904056, 11533005) and 973 project of China (2014CB744203) .

\end{acknowledgements}

\begin{figure}
\plotone{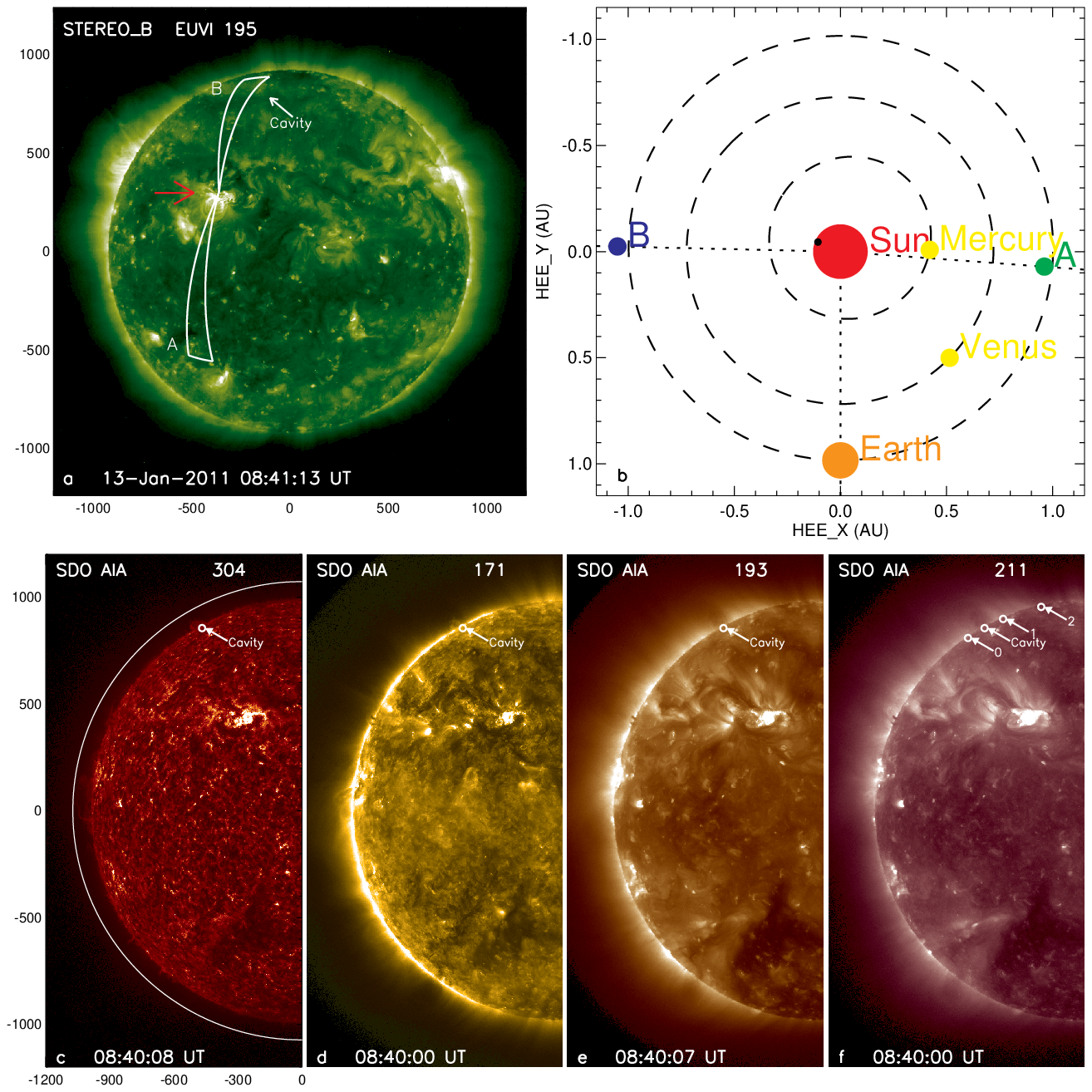}\caption{
Coronal environment before the eruption in EUVI-B (a) and AIA (c)---(f), as well as positions of the source AR (black solid circle on the Sun) and the \emph{STEREO} twin spacecraft on the date of the event (b). In EUVI-B, a dark channel on the disk is identified, which corresponds to a coronal cavity over the limb seen in AIA. The two sectors (labeled as ``A" and ``B") extending from the AR core (pointed out by the red arrow) in EUVI-B and the semi-circle above the limb in AIA are selected to construct time--distance/PA stack plots for the EUV wave, within which the kinematics of the wave front is studied. Note that in panel (f) the digits ``1'', ``2", and ``3'' on the limb are located at the southern, northern boundaries of the cavity, and the boundary of the northern polar coronal hole, corresponding to PAs of 56$^{\circ}$, 67$^{\circ}$, and 78$^{\circ}$, respectively.
}
\end{figure}

\begin{figure}
\plotone{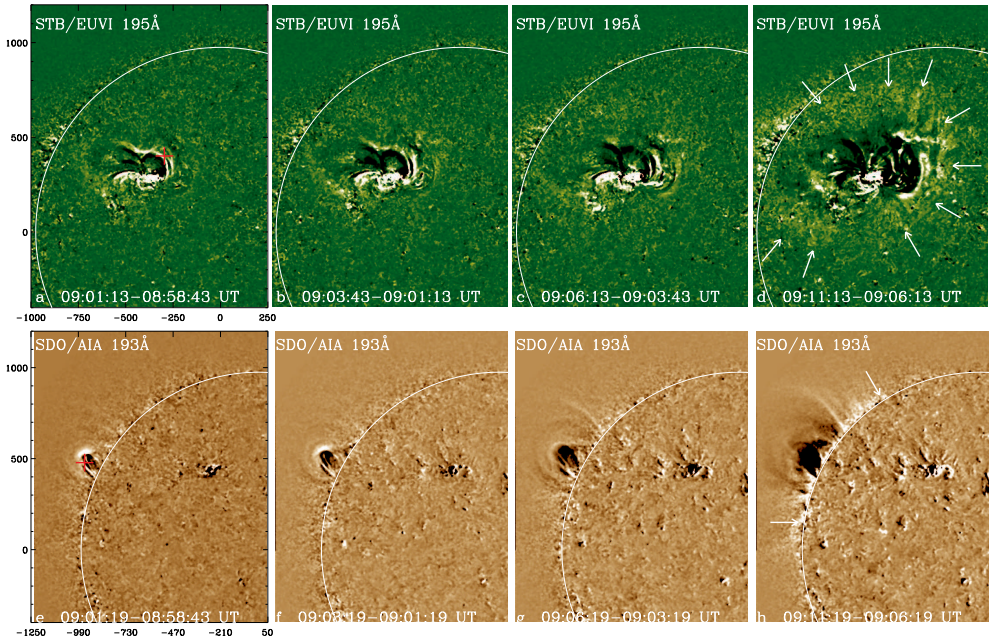}\caption{
Snapshots of the early stage evolution of the eruption and the associated EUV wave in EUVI-B 195 {\AA} (top) and AIA 193 {\AA} (bottom), which are extracted from the online animations. The plus signs outline the same feature in an erupting loop seen in different perspectives by using the triangulation method. The arrows point to the wave front when it is fully developed. 
}
\end{figure}

\begin{figure}
\epsscale{0.9}
\plotone{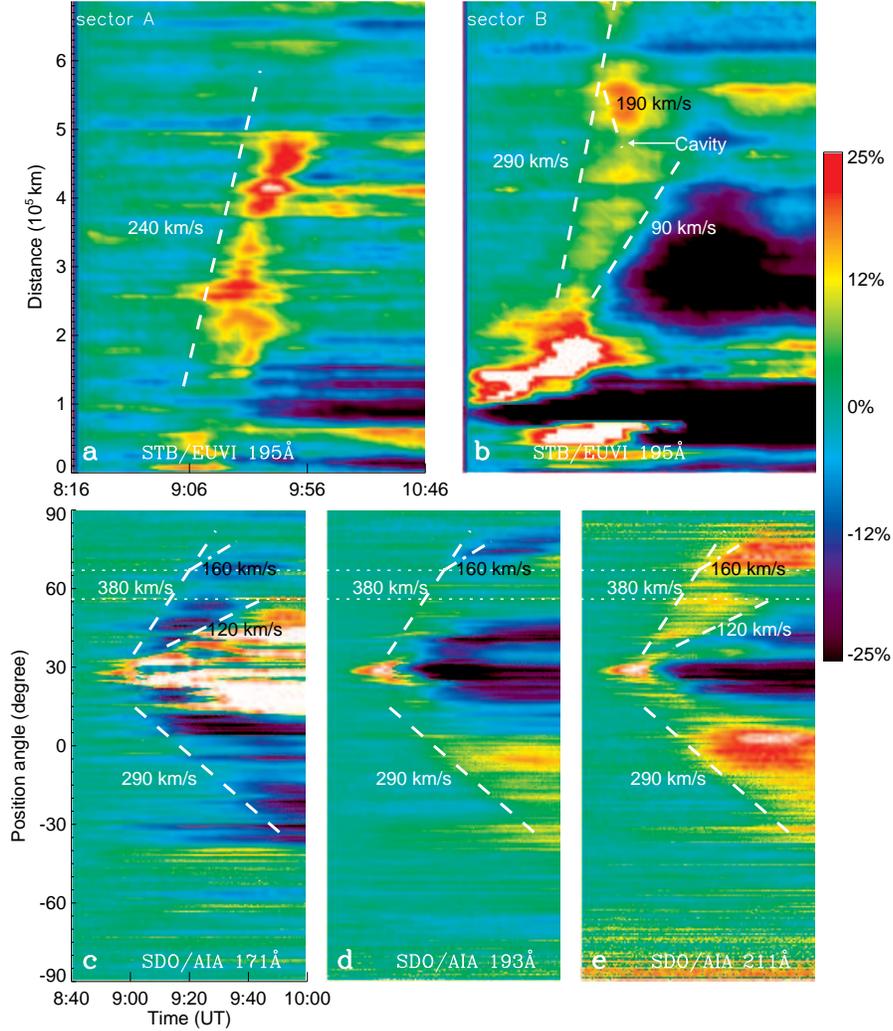} \caption{
Time--distance/PA stack plots for the two sectors in EUVI-B 195 {\AA} (top) and the semi-circle above the limb in AIA 171, 193, and 211 {\AA} (bottom). The plots are shown in base ratio, with the intensities color-coded according to the color bar on the right. Moving structures on the wave fronts are outlined by the dashed lines, and their propagation velocities are determined by applying linear fits. In panels (d)--(f) the two horizontal dotted lines correspond to the southern (56$^{\circ}$) and northern (67$^{\circ}$) boundaries of the coronal cavity, respectively.
}
\end{figure}

\begin{figure}
\plotone{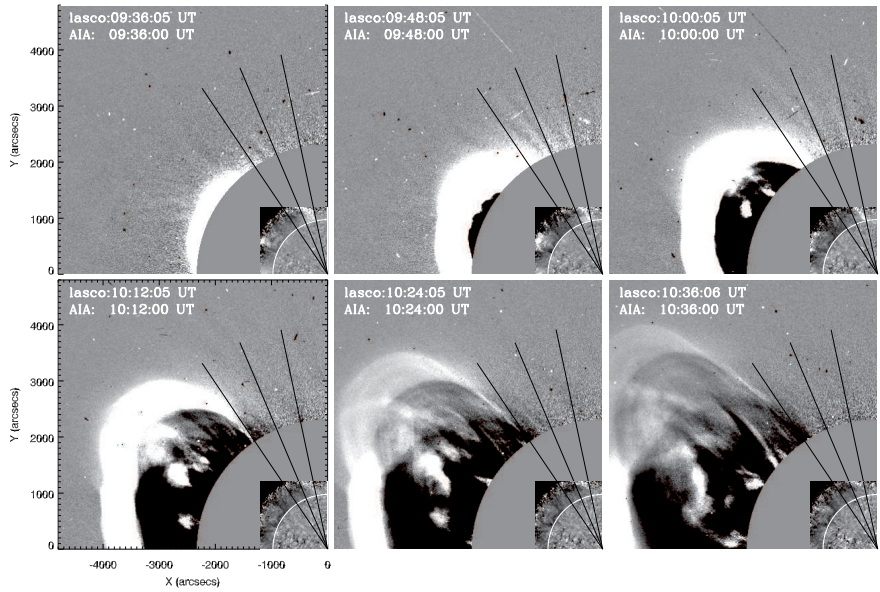} \caption{
Composite images of LASCO C2 and AIA in 211 {\AA},  demonstrating the evolution of the CME and the EUV wave. In a clockwise direction, the three radial lines are drawn along PAs of 56$^{\circ}$, 67$^{\circ}$, and 78$^{\circ}$, respectively.
}
\end{figure}

\begin{figure}
\plotone{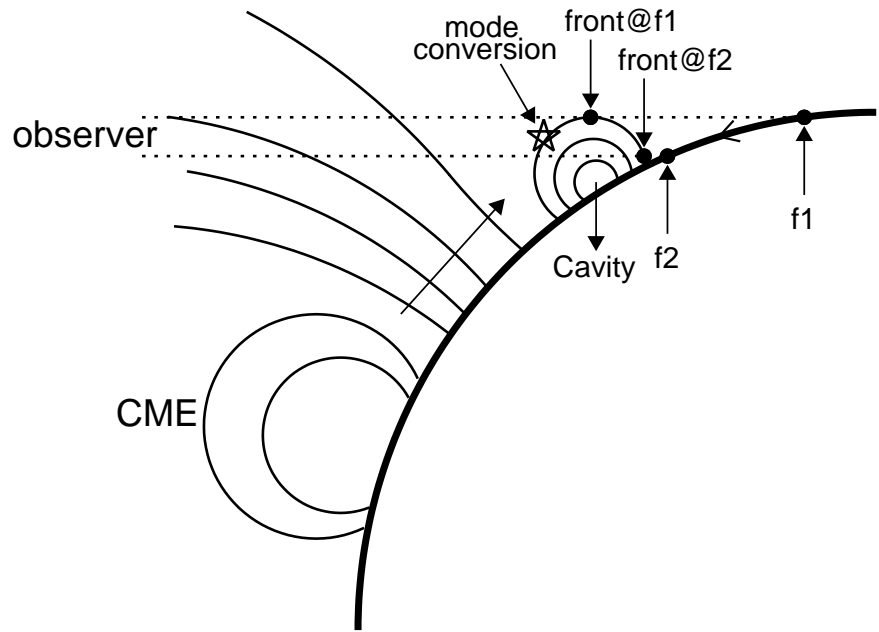} \caption{
Schematic plot showing the interaction of the leading edge (a fast-mode wave) and the cavity and the consequent mode conversion process.
}
\end{figure}

\end{document}